\documentstyle[12pt]{article}

\begin{document}

\setcounter{page}{0}

\begin{titlepage}

\title{Spin glass freezing in Kondo lattice compounds}

\author{Alba Theumann\\
	Instituto de F\'{\i}sica  - UFRGS\\
      Av. Bento Gon\c{c}alves 9500\\
      91501-970 Porto Alegre, RS, Brazil\\[4mm]
	B. Coqblin\\
	Laboratoire de Physique des Solides\\
	Universit\'e Paris - Sud\\
      91405 Orsay Cedex, France\\[4mm]
	S. G. Magalh\~aes and A. A. Schmidt\\
         Departamento de Matem\'atica\\
         Universidade Federal de Santa Maria\\
         97119-900 Santa Maria, RS, Brazil\\}

\date{}

\maketitle

\thispagestyle{empty}

\newpage

\begin{abstract}
It is presented a theory that describes  a spin glass 
phase at finite temperatures in Kondo lattice systems
with an additional RKKY interaction represented by
long range, random couplings among localized spins
like in the Sherrington- Kirkpatrick (SK) spin glass 
model. The problem is studied within the functional 
integral formalism
where the spin operators are represented by bilinear combinations
of fermionic (anticommuting) Grassmann variables.The Kondo and 
spin glass transitions are both described with the mean field 
like static ansatz that reproduces good results in the two well
known limits. At high temperatures and low values of the
Kondo coupling there is a paramagnetic
(disordered) phase with vanishing Kondo and spin glass
 order parameters. By lowering the temperature  
 a second order transition line is found at $T_{SG}$
to a spin glass phase.
For larger values of the Kondo coupling  there is a second order transition 
line at roughly $ T_{k}$ to a Kondo ordered state.
For $T<T_{SG}$ the transition between the Kondo 
and spin glass phases becomes first order.\\

\normalsize

\end{abstract}
\vspace*{0.5cm}

\end{titlepage}

%\end{document}
%\renewcommand{\baselinestretch}{1.5}

\setcounter{page}{1}

\section*{1.\ Introduction}

The antiferromagnetic s-f exchange coupling of conduction electrons to
 localized spins in heavy fermion rare-earth systems is responsible for 
two competing effects: the screening of the localized moments
 due to the Kondo effect and the Ruderman-Kittel-Kasuya-Yosida (RKKY)
interaction among magnetic impurities which may induce a long-range
magnetic (antiferromagnetic or ferromagnetic) ordering or eventually a 
spin glass magnetic ordering. The Doniach diagram\cite{1} gives a good 
description of 
this competition: the Neel temperature $ T_{N}$is firstly increasing  
with increasing 
the absolute value of the exchange interaction
constant $J_{k}$, then it is passing through a maximum and finally it 
tends to zero at the "quantum critical point" (QCP). Such a decrease of 
$ T_{N}$ down to the QCP has been observed in many Cerium compounds, 
such as$ CeAl_{2}, CeAg$ or $CeRh_{2}Si_{2}$, under pressure. 
Above the QCP, there exists a very strong heavy fermion 
character, but several possible behaviours, i.e. the classical 
Fermi liquid one with eventually
a reduced Kondo temperature\cite{2,3} or different Non-Fermi-Liquid (NFL)
ones, have been observed in Cerium or Ytterbium compounds\cite{4,5}.

  But, in the case of disordered Cerium alloys, the disorder can yield
a Spin Glass (SG) phase in addition to the NFL behaviour at low 
temperatures around the QCP.
The magnetic phase diagram of $CeNi_{1-x}Cu_{x}$ has been extensively 
studied\cite{6,7}. 
CeCu is antiferromagnetic below 3.5 K and CeNi is a
non magnetic compound with an intermediate valence.The low temperature 
antiferromagnetic phase changes, around $x=0.8$, to a ferromagnetic one 
which finally disappears around $x=0.2$. At higher temperatures , 
a Spin Glass state is deduced from all measured bulk 
properties, such as for example the ac susceptibility; for example, 
for $x=0.6$, the SG state
exists between 2 K and the Curie temperature $T_{c}=1.1 K$. At $x=0.2$,
there exists below 6 K only a SG state which transforms to the 
intermediate valence CeNi, 
after passing probably through a Kondo lattice system. Thus,
in the $ CeNi_{1-x}Cu_{x}$ system, there appears a SG phase at 
finite temperatures above the ferromagnetic order and finally a SG-Kondo 
transition occurs for small $x$ values.

  Another disordered Cerium system, namely $CeCoGe_{3-x}Si_{x}$ alloys,
has been also studied for different $x$ values, by different experimental techniques 
including muon spin relaxation\cite{8,9}. The compound 
$CeCoGe_{3}$ is antiferromagnetic below 21 K and $CeCoSi_{3}$ is an
intermediate valence compound ; the QCP of these alloys lies around 
$x=1.5$. The muon spin relaxation experiments in the quantum critical 
region ($x=1.1$ to $1.5$) show that a fraction of Ce ions experience 
random f-f indirect exchange interactions, which causes frustration 
of some Ce spins in the system. So, for $x=1.1$ and $x=1.2$, frustrated moments of 
Cerium freeze like in a spin glass while the other Ce moments form a disordered 
antiferromagnetic state. Thus, near the QCP, a spin glass state 
can exist in these alloys, in addition to the observed NFL behaviour.
So, a striking novel behaviour, i.e. the appearance of a spin glass 
state at finite temperatures in some disordered Cerium alloys, has been observed and 
a SG-Kondo transition is developing with increasing $x$ 
around the QCP.
 
The aim of our paper is to present a theoretical model that describes the 
spin glass-Kondo phase transitions, and that we do by studying a system 
Hamiltonian that couples  the localized spins of 
a Kondo lattice with an additional long range random interaction, 
like in the Sherrington- Kirkpatrick spin glass model \cite{10}. 
A similar 
Hamiltonian has been considered in \cite{11} to describe NFL behaviour 
and a QCP in some heavy fermions compounds, although the relevant 
approximations differ in this work and ours. The authors in \cite{11} 
are primarily interested in the description of the QCP at $T=0$, then 
they solve first for the Kondo effect by decoupling the conduction 
electrons bath into independent conduction electron "reservoirs", 
with  no communication between the reservoirs at different sites.
 
During the course of our work, another paper using the same 
Hamiltonian has been proposed\cite{12}. The representation
of Popov and Fedotov eliminates the unwanted spin states but the 
approximations involved are in fact different from those used in our work.
They study essentially the spin glass state and they finally obtain a 
second order SG transition with a  transition temperature depressed by the Kondo effect, in second order perturbation theory.

In the present paper we take a different approach: the localized 
spins of the Kondo lattice will be effectively immersed in a common 
bath of conduction electrons and the Kondo effect will be studied in 
a quantum static approximation that is basically equivalent to the 
mean field decoupling scheme  \cite{3,13}.
The spin operators are represented by bilinear combinations of 
fermionic creation and destruction operators, for the localized 
f-electrons, and the spin glass transition will be studied within 
the static approximation. This deserves some special discussion. In 
the Ising quantum spin glass (QSG) model  \cite{14}, the spin operator 
$S_{i}^{z}$ commutes with the particle number operator $n_{is}$  
 and the static Ansatz gives the exact answer, as the problem is 
essentially classic. When we add to the fermionic Ising QSG a s-d 
exchange coupling of the localized spins to the conduction band 
electrons the problem ceases to be exactly soluble and the static 
Ansatz is just an approximation, that we consider justified to 
describe a transition at finite temperature. It has been 
shown in \cite{15} that the exact numerical solution of Bray and Moore's 
equations  \cite{16} gives for the spin-spin correlation function 
$Q(\tau)$ roughly its constant classical value at finite temperature, 
what justifies the use of the static Ansatz of \cite{16} at not very low
temperatures in the Heisenberg spin glass. 

We consider then the static Ansatz  which corresponds
to an approximation similar to mean 
field theory, where by neglecting time fluctuations we can provide a 
description of the phase transitions occurring at finite temperature.

We use functional integral techniques where the spin operators are 
represented by bilinear combinations of fermionic (anticommuting) 
Grassmann fields. As we show in the next section, this method is 
ideally suited to describe a Kondo lattice transition, and it has been
recently applied by two of us to the study of fermionic Ising spin 
glasses with local BCS pairing  \cite{17}. Recent work  \cite{18} also 
showed the existence of several characteristic temperatures in the 
Ising fermionic model, with the de Almeida-Thouless instability  
\cite{19} occurring at a temperature lower than the spin glass 
transition temperature.

This paper is organized as follows: in Sect.2 we describe the model 
and relevant results, we reserve Sect.3 for discussions and conclusion, 
while the detailed mathematical calculations are left for the Appendix.

\section*{2.\ The model and results}

We consider a Kondo lattice system with localized spins 
$\vec{S_{i}}$ at sites $i=1 \cdots N$, coupled to the electrons of 
the conduction band via a s-d exchange interaction. It is necessary 
to introduce explicitly the resultant RKKY interaction by means of a 
random, long range coupling among localized spins like in the 
Sherrington Kirkpatrick (SK) model for a spin glass. To describe the 
Kondo effect in a mean-field-like theory it is sufficient to keep only 
the spin-flip terms  \cite{3} in the exchange Hamiltonian, while the 
spin glass interaction is represented by the quantum Ising Hamiltonian 
where only interact the z-components of the localized spins  
\cite{11,14,17}.

The Hamiltonian of the model is:

\begin{eqnarray}
{\cal H}-\mu_{c}N_{c}-\mu_{f}N_{f}= 
{\cal H}_{k}-\mu_{c}N_{c}-\mu_{f}N_{f} + {\cal H}_{SG}
\label{2.1}
\end{eqnarray}

\begin{eqnarray}
{\cal H}_{k}-\mu_{c}N_{c}-\mu_{f}N_{f}= \displaystyle \sum_{k,\sigma}\epsilon_{k}n_{k\sigma}
+\epsilon_{0} 
\displaystyle \sum_{i,\sigma}n_{i\sigma}^{f}+J_{k}\displaystyle 
\sum_{i}[S_{fi}^{+}s_{ci}^{-}+S_{fi}^{-}s_{ci}^{+}]
\label{2.2}
\end{eqnarray}

\begin{eqnarray}
{\cal H}_{SG}=- \displaystyle \sum_{i,j}J_{ij}S_{fi}^{z}S_{fj}^{z}
\label{2.3}
\end{eqnarray}
\noindent
where $J_{k} > 0 $,

\begin{eqnarray}
{S_{fi}^{+}}=f_{i\uparrow}^{\dagger}f_{i\downarrow}\hspace{0.8cm};
\hspace{0.8cm}s_{ci}^{+}=d_{i\uparrow}^{\dagger}
d_{i\downarrow}\nonumber\\
{S_{fi}^{-}}=f_{i\downarrow}^{\dagger}f_{i\uparrow}\hspace{0.8cm};
\hspace{0.8cm}s_{ci}^{-}=d_{i\downarrow}^{\dagger}d_{i\uparrow}\\
{S_{fi}^{z}}=\frac{1}{2}[f_{i\uparrow}^{\dagger}f_{i\uparrow}-f_{i\downarrow}^{\dagger}f_{i\downarrow}]\nonumber
\label{2.4}
\end{eqnarray}
and $f_{i\sigma}^{+},f_{i\sigma}(d_{i\sigma}^{+},d_{i\sigma})$ are 
creation and destruction operators for electrons with spin projection 
$\sigma=\uparrow $ or $\downarrow$ in the localized (conduction) band, 
that satisfy  the standard fermion anticommutation rules. We also have 
$n_{k\sigma}=d_{k\sigma}^{+}d_{k\sigma}$ where:
\begin{eqnarray}
d_{k\sigma}=\frac{1}{\sqrt{N}}\displaystyle 
\sum_{i}e^{i\vec{k}\cdot\vec{R_{i}}}d_{i\sigma}\nonumber\\
d_{k\sigma}^{+}=\frac{1}{\sqrt{N}}\displaystyle 
\sum_{i}e^{-i\vec{k}\cdot\vec{R_{i}}}d_{i\sigma}^{+}
\label{2.5}
\end{eqnarray}
\noindent
The energies $\epsilon_{0}(\epsilon_{k})$ are referred to the 
chemical potentials $\mu_{f}(\mu_{c})$, respectively.

The coupling $J_{ij}$ in eq.(\ref{2.3}) is an independent random 
variable with the gaussian distribution

\begin{eqnarray}
P(J_{ij})=e^{-J_{ij}^{2}\frac{N}{32J^{2}}}\sqrt{\frac{N}{32\pi J^{2}}}          
\label{2.6}
\end{eqnarray}

Functional integration techniques have proved to be a suitable 
approach to describe phase transitions in disordered quantum 
mechanical many-particle systems  \cite{20}. The static approximation 
within this formulation consists in neglecting time fluctuations of 
the order parameter, and when it is combined with the neglect of space 
fluctuations it leads to the usual Hartree-Fock, mean field like 
approximation. When dealing with the Hamiltonian in 
Eq. (\ref{2.1})-Eq.(\ref{2.3}), we notice that in the limiting case 
$J_{k}=0$ we obtain a pure quantum Ising spin glass where the static 
approximation gives the exact result  \cite{14,18}, while for $J=0$
we recover the mean field approximation that has been used succesfully 
to describe the Kondo lattice \cite{3,13}. Then we consider that the use 
of the static approximation has an interpolation character and 
will provide reliable results to describe 
critical behaviour at finite temperature  \cite{15} in systems that do not
present a quantum critical point.

In the Lagrangian formulation  \cite{17,20} the partition function is 
expressed as

\begin{eqnarray}
Z=\displaystyle \int D(\varphi^{\dag}\varphi)(\psi^{\dag}\psi)e^{A}
\label{2.7}
\end{eqnarray}

\noindent
where the action $A$ is given by

\begin{eqnarray}
A=\displaystyle\sum_{i,\sigma}\displaystyle\int_{0}^{\beta}d\tau
\{(\frac{\partial}{\partial\tau}
\varphi_{i\sigma}^{\dag})\varphi_{i\sigma}+(\frac{\partial}
{\partial\tau}\psi_{i\sigma}^{\dag})\psi_{i\sigma}\}-\displaystyle 
\int_{0}^{\beta}{\cal H}(\tau)d\tau
\label{2.8}
\end{eqnarray}

In both expressions $\varphi_{i\sigma}(\tau)$ and 
$\psi_{i\sigma}(\tau)$ are anticommuting, complex Grassmann 
variables associated to the conduction and localized electrons fields,
respectively, while $\tau$ is an imaginary time and $\beta$ the 
inverse absolute temperature.

We show in the Appendix that in the static, mean field like approximation 
the action A may be written:

\begin{eqnarray}
A=A_{0}+A_{K}+A_{SG}
\label{2.9}
\end{eqnarray}

with

\begin{eqnarray}
A_{0}=\displaystyle\sum_{\omega\sigma}\displaystyle
\sum_{i,j}[(i\omega-\beta\epsilon_{0})\delta_{ij}\psi_{i\sigma}^{\dag}
(\omega)\psi_{i\sigma}(\omega)+(i\omega\delta_{ij}-\beta t_{ij})
\varphi_{i\sigma}^{\dag}(\omega)\varphi_{j\sigma}(\omega)]
\label{2.10}
\end{eqnarray}

where from Eq.(\ref{A.II})

\begin{eqnarray}
A_{K}=\frac{\beta J_{k}}{N}\displaystyle\sum_{\sigma} 
[\displaystyle\sum_{i\omega}\psi_{i,\sigma}^{\dag}
(\omega)\varphi_{i\sigma}(\omega)][\displaystyle\sum_{i,\omega}
\varphi_{i-\sigma}^{\dag}\psi_{i-\sigma}(\omega)]
\label{2.11}
\end{eqnarray}

\begin{eqnarray}
A_{SG}=\displaystyle\sum_{i,j} J_{ij}S_{i}^{z}S_{j}^{z}
\label{2.12}
\end{eqnarray}

and in the static approximation \cite{14,17}

\begin{eqnarray}
S_{i}^{z}=\frac{1}{2}\displaystyle\sum_{\sigma}\displaystyle
\sum_{\omega}\psi_{i\sigma}^{\dag}(\omega)\psi_{i\sigma}(\omega)
\label{2.13}
\end{eqnarray}
\noindent
The sums are over fermion Matsubara frequencies $\omega=(2n+1)\pi$.

The Kondo order is described by the complex order parameter

\begin{eqnarray}
\lambda_{\sigma}^{\dag}=\frac{1}{N}\displaystyle
\sum_{i,\omega}\langle\psi_{i\sigma}^{\dag}(\omega)
\varphi_{i\sigma}(\omega)\rangle
\nonumber\\
\lambda_{\sigma}=\frac{1}{N}\displaystyle
\sum_{i,\omega}\langle\varphi_{i\sigma}^{\dag}(\omega)
\psi_{i\sigma}(\omega)\rangle
\label{2.14}
\end{eqnarray}
that in a mean field theory \cite {3,13} describes the correlations 
$\lambda_{\sigma}^{\dag}=\langle f_{i\sigma}^{\dag}d_{i\sigma}\rangle$ 
and  $\lambda_{\sigma}=\langle d_{i\sigma}^{\dag}f_{i\sigma}\rangle$
 - Complex conjugation of Grassmann variables is defined through the 
transposition rule \cite{20} $(\psi^{\dag}\varphi)^{\dag}=
\varphi^{\dag}\psi$.

We show in the Appendix that standard manipulations give for the 
averaged free energy within a replica symmetric theory:
\begin{eqnarray}
\beta F=2\beta J_{k}\lambda^{2}+\frac{1}{2}\beta^{2}J^{2}
(\overline{\chi}^{2}+2q\overline{\chi})
-\beta\Omega \label{2.15}
\end{eqnarray}
where
\begin{eqnarray}
\beta\Omega=\lim_{n \rightarrow 0}\frac{1}{Nn}\{\displaystyle
\int_{-\infty}^{+\infty}\displaystyle \prod_{j}^{N}Dz_{j}\displaystyle
 \prod_{\alpha}^{n}\displaystyle \int_{-\infty}^{+\infty}\displaystyle 
\prod_{j}D\varepsilon_{\alpha j}\exp{(\displaystyle 
\sum_{\omega ,\sigma}\ln{|G_{ij\sigma}^{-1}(\omega)|})}-1\}
\label{2.16}
\end{eqnarray}
and the order parameters $q, \overline{\chi}$, and $\lambda$ must be 
taken at their saddle point value. Here $q$ is the SG order parameter 
\cite {14,17} and the static susceptibility is $\chi=\beta \overline
{\chi}$. We use the notation 
$Dx=\frac{dx}{\sqrt{2\pi}}e^{-\frac{1}{2}x^2}$.

The function $G_{ij\sigma\alpha}(\omega)$ in Eq.(\ref {2.16}) is the time 
Fourier transform of the Green's function $G_{ij\sigma}(\tau)=
i\langle T f_{i\sigma}(\tau)f_{j\sigma}^{\dag}(0)\rangle$ for  the 
localized electrons in the presence of random fields  $z_{j}$ and 
$\varepsilon_{\alpha j}$ at every site, and from Eq.(\ref{A.XXIII})  satisfies the equation
\begin{eqnarray}
G_{ij\sigma\alpha}^{-1}(\omega)=[i\omega-\beta\epsilon_{0}-
\sigma h_{j\alpha}]\delta_{ij}-
\beta^{2}J_{k}^{2}\lambda^{2}\gamma_{ij}(\omega)
\label{2.17}
\end{eqnarray}
where
\begin{equation}
h_{j\alpha}=\sqrt{2q}\beta Jz_{j}+\sqrt{2\overline{\chi}}\beta 
J\varepsilon_{\alpha j}
\label{2.18}
\end{equation}
while $\gamma_{ij}(\omega)$ is the time Fourier transform of the 
conduction electron Green'\nolinebreak s function  
$\gamma_{ij}(\tau)=i\langle Td_{i\sigma}(\tau)d_{i\sigma}^
{\dag}(0)\rangle$ and is given by
\begin{equation}
\gamma_{ij}^{-1}=[i\omega-\beta\mu_{c}]\delta_{ij}-\beta t_{ij}
\label{2.19}
\end{equation}

  We  obtained in Eq.\ (\ref {2.17})  the Green's function for the f-electrons 
in a Kondo lattice \cite{13}, but now in the presence of a random field 
$h_j$ at every site  that prevents us from proceeding with the calculation. 
In the pure SG limit $J_{k}=0$ the Green's function 
in Eq.\ (\ref{2.17})is local and the integrals in Eq.\ (\ref{2.16})
 reduce to a one site problem, while in the Kondo limit $J=0$
the random fields vanish and the integrals separate in reciprocal space. 
We adopt here a decoupling approximation that is reminiscent of the model  with 
independent "reservoirs" of electrons considered in \cite{11}.
We replace the Green's function 
$G_{ij\sigma\alpha}(\omega,\{h_{1\alpha}...h_{j\alpha}...h_{N\alpha}\})$ 
by the Green's functions $\Gamma_{\mu\nu\sigma}(\omega,h_{j\alpha}), j=1...N$, of $N$ independent Kondo lattices, 
each one with a "uniform" field $h_{j\alpha}$ at every site $\mu, \nu$,
 by means of the approximation:
\begin{equation}
\ln{|G_{ij\sigma\alpha}^{-1}(\omega,\{h_{1}...h_{N}\})
|\approx\frac{1}{N}\displaystyle\sum_{j}\ln{|\Gamma_{\mu\nu\sigma}^{-1} 
(\omega ,h_{j\alpha})|}}
\label{2.20}
\end{equation}
where $\Gamma_{\mu\nu\sigma}(\omega,h_{j\alpha})$ is the f-electron 
Green's function for a fictitous Kondo lattice that has a uniform field
 $h_{j\alpha}$ at every site $\mu,\nu$ and satisfies the equation:
\begin{eqnarray}
\Gamma_{\mu\nu\sigma}^{-1}(\omega ,h_{j})=[i\omega -\beta\epsilon_{0}-
\sigma h_{j\alpha}]\delta_{\mu\nu}-\beta^{2}J_{k}^{2}\lambda^{2}
\gamma_{\mu\nu}
\label{2.21}
\end{eqnarray}
where, from Eq.\ (\ref {2.19}):
\begin{equation}
\gamma_{\mu\nu}(\omega)=\frac{1}{N}\displaystyle \sum_{k}\frac{1}
{i\omega-\beta\epsilon_{k}}e^{i\vec{k}\cdot\vec{R}_{\mu\nu}} 
\label{2.22}
\end{equation}
Now Eq.\ (\ref {2.21}) may be easily solved by a Fourier transformation
with the result:
\begin{eqnarray}
\ln{|\Gamma_{\mu\nu\sigma}^{-1}(\omega ,h_{j\alpha})|}=\displaystyle
\sum_{\vec{k}}\ln{[\Gamma_{\vec{k}\sigma}^{-1}(\omega ,h_{j\alpha})]}
\label{2.23}
\end{eqnarray}
where
\begin{eqnarray}
\Gamma_{\vec{k}\sigma}^{-1}(\omega ,h_{j\alpha})=
[i\omega-\beta\epsilon_{0}-\sigma h_{j\alpha}]-
\beta^{2}J_{k}^{2}\lambda^{2}\frac{1}{i\omega-\beta\epsilon_{k}}.
\label{2.24}
\end{eqnarray}

We may now introduce Eq.\ (\ref {2.20}) and Eq.\ (\ref {2.23}) in 
Eq.\ (\ref {2.16}), the integrals over the fields separate and we obtain 

\begin{equation}
\beta\Omega=\displaystyle\int_{-\infty}^{+\infty}Dz\ln{\{\displaystyle
\int_{-\infty}^{+\infty}D\varepsilon\exp{(\displaystyle\sum_{\sigma}
\frac{1}{N}\displaystyle\sum_{\vec{k}}S_{\sigma}(\vec{k},h))}}\}
\label{2.25}
\end{equation}
with
\begin{equation}
S_{\sigma}(\vec{k},h)=\displaystyle\sum_{\omega}
\ln{[\Gamma_{\vec{k},\sigma}^{-1}(\omega,h)]}
\label{2.26}
\end{equation}
and $h$ is  given in Eq.\ (\ref {2.18}), with $z$ and $\varepsilon$ 
in place of $z_{j}$ and $\varepsilon_{j\alpha}$.

The sum over the fermion frequencies is performed in the standard way 
by integrating  in the complex plane\cite{20}, with the result:
\begin{equation}
S_{\sigma}(\vec{k},h)=\ln{[(1+e^{-\omega_{\sigma^{+}}})
(1+e^{-\omega_{\sigma^{-}}})]}
\label{2.27}
\end{equation}
where 
\begin{equation}
\omega_{\sigma^{\pm}}=\frac{1}{2}[\beta\epsilon_{k}+\sigma h]\pm
\{\frac{1}{4}(\beta\epsilon_{k}-\sigma h)^{2}+
(\beta J_{k}\lambda)^{2}\}^{\frac{1}{2}}.
\label{2.28}
\end{equation}
We consider $\epsilon_{0}=0$ that corresponds to an average 
occupation $\langle n_{F}\rangle=1$, per site.

Replacing sums by integrals , in the approximation of a constant 
density of states for the conduction band electrons, $ \rho(\epsilon)=
\rho=\frac{1}{2D}$ for $-D<\epsilon <D$, we obtain from 
Eq.\ (\ref {2.25}) to Eq.\ (\ref {2.27}) the final expression for the 
free energy in Eq.\ (\ref {2.15}):
\begin{eqnarray}
\beta F=2\beta J_{k}\lambda^{2}+\frac{1}{2}\beta^{2}J^{2}
(\overline{\chi}^{2}+2q\overline{\chi})-
\displaystyle\int_{-\infty}^{+\infty}Dz\ln{\{\displaystyle
\int_{-\infty}^{+\infty}D\varepsilon e^{E(h)}\}}
\label{2.29}
\end{eqnarray}
with
\begin{eqnarray}
E(h)=\frac{1}{\beta D}\displaystyle\int_{-\beta D}^{+\beta D}dx
\ln{\{\cosh{\frac{(x+h)}{2}}+\cosh{(\sqrt{\Delta})}\}},
\label{2.30}
\end{eqnarray}
\begin{eqnarray}
\Delta=\frac{1}{4}(x-h)^2+(\beta J_{k}\lambda)^2
\label{2.31}
\end{eqnarray}
and from Eq.\ (\ref {2.18}) we have 
$h=\beta J[\sqrt{2q}z+\sqrt{2\overline{\chi}}\varepsilon]$. 
The saddle point equations for the SG order parameters are:
\begin{eqnarray}
q=\displaystyle\int_{-\infty}^{+\infty}Dz
\{\frac{\int D\varepsilon e^{E}\frac{\partial E}{\partial h}}
{\int D\varepsilon e^{E}}\}^2
\label{2.32}
\end{eqnarray}
\begin{eqnarray}
\overline{\chi}=\displaystyle\int_{-\infty}^{+\infty}Dz\frac{1}
{\int D\varepsilon e^{E}}\{\int D\varepsilon\frac{\partial}
{\partial h}[e^{E}\frac{\partial E}{\partial h}]\}
\label{2.33}
\end{eqnarray}

while we obtain for the Kondo order parameter $\lambda$

\begin{eqnarray}
4\beta J_{k}\lambda\{1-\frac{\beta J_{k}}{4}\displaystyle
\int_{-\infty}^{+\infty}Dz\frac{1}{\int D\varepsilon e^{E}}\int 
D\varepsilon e^E \nonumber\\
\times\frac{1}{\beta D}\displaystyle\int_{-\beta D}^{+\beta D}dx\frac{1}
{\cosh{(\frac{x+h}{2})}+\cosh{(\sqrt{\Delta}})}
(\frac{\sinh{(\sqrt{\Delta}})}{\sqrt{\Delta}})\}=0
\label{2.34}
\end{eqnarray}

The numerical solution of the saddle point equations as a
function of $\frac{T}{J}$ and $\frac{J_{k}}{J}$ provides us 
with the phase diagram in Fig.1, that we discuss in the next section.

\section*{3.\ Conclusions}
We study in this paper the phase transitions in a system 
represented by a Hamiltonian that couples the localized 
spins of a Kondo lattice \cite{3,13}with random, long range 
interactions, like in the SK model for a spin glass\cite{10}.\\

Using functional integrals techniques and a static, replica 
symmetric Ansatz for the Kondo and spin glass order parameters,
we derive a mean field expression for the free energy and the
saddle point equations for the order parameters. The Kondo and 
spin glass transitions are both described with the mean field 
like static ansatz that reproduces good results in two well
known limits: when $J_{k}=0$ we recover the exact solution for 
the quantum Ising spin glass \cite{14,18} while for $J=0$ we recover 
the mean field results for the Kondo lattice \cite{3,13}.The use of the 
static ansatz is justified at finite temperatures\cite{15}. Numerical
solution of the saddle-point equations allow us to draw the magnetic
phase diagram in the $J_{k} vs T$ plane, for fixed value of $J$,
presented in Fig.1.\\

  Figure 1 shows three different phases. At high temperatures, the 
"normal" phase is paramagnetic with vanishing Kondo and spin glass
order parameters, i.e. $\lambda= q= 0$. When temperature is lowered, 
for not too large values of the ratio $J_{K}/J$, a second-order transition
 line is found at $T= T_{SG}$ to a spin glass phase with $q >0$ and 
$\lambda= 0$. Finally, for large values of the ratio $J_{K}/J$,  
we recover the "Kondo" phase with a non-zero $\lambda$ value 
and $q=0$: the transition line from the paramagnetic phase to the Kondo phase for 
temperatures larger than $T_{SG}$ is a second-order one and 
occurs at a temperature very close to the one-impurity Kondo temperature $T_{K}$.
On the other hand, the transition line from the spin-glass phase to the Kondo phase, 
for temperatures smaller than  $T_{SG}$, is a first-order 
one and it ends at $J_{K}^{c}$ at $T=0$. When the temperature is lowered,
the transition temperature does not vary very much with the value of
$J_{K}/J$; the separation between the spin-glass and the Kondo phases departs completely 
from the behaviour of $T_{K}$ and looks like the separation between the magnetic and Kondo 
phases when these two phases are considered\cite{21}. We can also remark that we get here 
only "pure"
Kondo or SG phases and never a mixed SG-Kondo phase with the two order parameters 
different from zero; this result is probably connected to the approximations used 
here to treat the starting Hamiltonian.

  The diagram shown in figure 1 can explain the magnetic phase diagram observed 
above the Curie temperature for the $CeNi_{1-x}Cu_{x}$ \cite{6}
for small $x$ values when there is a transition from a spin-glass state 
to a Kondo state and then to the intermediate valence compound CeNi; however, 
there is no experimental information on the precise nature of the SG-Kondo 
transition and our model cannot be checked from that point of 
view. There is also probably a SG-Kondo transition in the 
$ CeCoGe_{3-x}Si_{x}$ alloys, but there the experimental situation 
is even more complicate than in the preceding case.
 An unsolved basic question concerns also the existence or not of a "mixed" 
SG-Kondo phase in Cerium
disordered alloys and this problem is worth of being studied experimentally in more detail. 
Thus, further experimental work is necessary, but our model yields a new 
striking point in the behaviour of heavy fermion disordered alloys in the 
vicinity of the quantum critical point.

\section*{Appendix }
%\begin{document}
%\appendix

We  present here a detailed derivation of the main equations of the 
paper. By introducing eqs. (\ref{2.1})-(\ref{2.3}) in eq. (\ref{2.8}) 
we obtain for the s-d exchange part of the action:
\begin{equation}
A_{K}=-\beta J_{k}\displaystyle\sum_{i}\displaystyle
\sum_{w,w',\Omega}\displaystyle\sum_{\sigma}\psi_{i
\sigma}^{\dagger}(w)\psi_{i-\sigma}(w'+\Omega)\varphi_{i-\sigma}^
{\dagger}(w')\varphi_{i\sigma}(w-
\Omega)
\label{A.I}
\end{equation}                                                                                                                                        
where $w=(2n+1)\pi$ and $\Omega=2n\pi$. In the mean field spirit we want 
to introduce the spatially uniform and static Kondo order parameter in 
eq. (\ref{2.14}), then we take $\Omega=0$, re-order the operators and 
separate the sites in eq.(\ref{A.I}) with the introduction of a 
$N^{-1}$ factor, which gives:
\begin{equation}
A_{K}\approx +\frac{\beta J_{k}}{N}\displaystyle\sum_{\sigma}
\displaystyle\sum_{i,w}\psi_{i
\sigma}^{\dagger}(w)\varphi_{i\sigma}(w)\displaystyle\sum_{j,w'}
\varphi_{j-\sigma}^{\dagger}(w')\psi_{j-\sigma}(w'),
\label{A.II}
\end{equation}                                                                                 that is eq.(\ref{2.11}).  

We find it convenient to introduce the Kondo order parameters in 
eq.(\ref{2.14}) by means of the identity
\begin{eqnarray}
e^{A_{k}}&=&\displaystyle\int_{-\infty}^{\infty}\displaystyle
\Pi_{\sigma} d\lambda_{\sigma}^{\dagger}d\lambda_{\sigma}
\displaystyle\Pi_{\sigma}\delta [\lambda_{\sigma}^{\dagger}N-
\displaystyle\sum_{j,w}\psi_{j\sigma}^{\dagger}(w)\varphi_{j\sigma}(w)]
\delta\nonumber\\
	 & &\times [\lambda_{\sigma}N-\displaystyle
\sum_{j,w}\varphi_{j\sigma}^{\dagger}(w)\psi_{j\sigma}(w)]
 e^{\beta J_{k}N[\lambda_{\uparrow}^{\dagger}\lambda_{\downarrow}+
\lambda_{\downarrow}^{\dagger}\lambda_{\uparrow}]},
\label{A.III}
\end{eqnarray}                                                                                 
and using the integral representation of the $\delta$-funtion:
\begin{equation}
\delta(x-x_{0})=\frac{1}{2\pi}\displaystyle\int_{-\infty}^{\infty}du e^{iu(x-x_{0})}
\label{A.IV}
\end{equation}
we may write the partition function by combining eq.(\ref{2.7}) to eq.(\ref{2.13}):

\begin{eqnarray}
Z&=&\displaystyle\int_{-\infty}^{\infty}\Pi_{\sigma} d\lambda_{\sigma}^{\dagger}
d\lambda_{\sigma}\displaystyle
\int_{-\infty}^{\infty}\Pi_{\sigma}\frac{du_{\sigma}dv_{\sigma}}{(2\pi)^{2}}\nonumber\\
 & &\times e^{N\displaystyle
\sum_{\sigma}\{\beta J_{k}\lambda_{\sigma}^{\dagger}\lambda_{-\sigma}-
i[u_{\sigma}\lambda_{\sigma}^{\dagger}+v_{\sigma}\lambda_{\sigma}]\}}Z_{eff}
\label{A.V}
\end{eqnarray}                                                                                 
where
\begin{eqnarray}
Z_{eff}&=&\int D(\psi^{\dagger}\psi)\int D(\varphi^{\dagger}\varphi )e^{A_{0}+A_{SG}}\nonumber\\
       & &\times e^{\displaystyle\sum_{\sigma}i[v_{\sigma}\displaystyle
\sum_{j,w}\varphi_{j\sigma}^{\dagger}(w)\psi_{j\sigma}(w)+u_{\sigma}\displaystyle
\sum_{j,w}\psi_{j\sigma}^{\dagger}(w)\varphi_{j\sigma}(w)]}
\label{A.VI}
\end{eqnarray}                                                               
The saddle point values of $\lambda_{\sigma}$, $\lambda_{\sigma}^{\dagger}$, $u_{\sigma}$, 
$u_{\sigma}^{\dagger}$ are otained by extremizing the exponent in eq.(\ref{A.V}):
\begin{eqnarray}
\beta J_{k}\lambda_{-\sigma}^{\dagger}=iv_{\sigma}\nonumber\\
\beta J_{k}\lambda_{-\sigma}=iu_{\sigma}\nonumber\\
\lambda_{\sigma}^{\dagger}=\frac{1}{N}\displaystyle
\sum_{j,w}\langle\psi_{j\sigma}^{\dagger}(w)\varphi_{i\sigma}(w)\rangle\nonumber\\
\lambda_{\sigma}=\frac{1}{N}\displaystyle
\sum_{j,w}\langle\varphi_{i\sigma}^{\dagger}(w)\psi_{j\sigma}(w)\rangle
\label{A.VII}
\end{eqnarray}
where $\langle ...\rangle=\int D(\psi^{\dagger}\psi) D(\varphi^{\dagger}\varphi)e^{A_{eff}}(...)$ 
from eq.(\ref{A.VI}). Introducing the saddle point values of eq.(\ref{A.VII}) into 
eq.(\ref{A.V}), we obtain:
\begin{equation}
Z=e^{-2N\beta J_{k}\lambda^{\dagger}\lambda}Z_{eff}.
\label{A.VIII}
\end{equation}

From eq.(\ref{2.10}), $Z_{eff}$ is now the integral of a  quadratic form in the 
$\varphi^{\dagger},\varphi$ variables, that can be integrated out to give:
\begin{equation}
Z_{eff}=Z_{d}^{0}Z_{SG}
\label{A.IX}
\end{equation}
where $Z_{d}^{0}$ is the partition function of the free conducting electrons:
\begin{equation}
\ln{(Z_{d}^{0})}=2\displaystyle\sum_{w}\log{|\gamma_{ij}^{-1}(w)|}
\label{A.X},
\end{equation}
and                    
\begin{equation}
\gamma_{ij}^{-1}(w)=iw\delta_{ij}-\beta t_{ij}
\label{A.XI},
\end{equation}
is the inverse  Green's function for the d-electrons. The quantity $Z_{SG}$ in eq.(\ref{A.IX}) 
is the partition function for the localized f-electrons:                                           \begin{equation}
Z_{SG}=\int D(\psi^{\dagger}\psi)e^{\displaystyle
\sum_{w\sigma}\sum_{i,j}g_{ij}^{-1}(w)\psi_{i\sigma}^{\dagger}(w)\psi_{j\sigma}(w)+A_{SG}}
\label{A.XII}
\end{equation}
where the inverse  Green's function for the localized, non-interacting f-electrons is now 
modified by the Kondo interaction:
\begin{equation}
g_{ij}^{-1}(w)=(iw-\beta\varepsilon_{0})\delta{ij}-\beta^{2}J_{k}^{2}\lambda^{\dagger}
\lambda\gamma_{ij}(w)
\label{A.XIII}
\end{equation}
and $A_{SG}$ is given in eq.(\ref{2.12}).

The interesting part of the free energy is given by:
\begin{equation}
\beta F=-\frac{1}{N}\langle\langle \ln{(\frac{Z}{Z_{d}^{0}})}\rangle\rangle
\label{A.XIV}
\end{equation}
where the double bracket indicates a configurational average over the random 
variables $J_{ij}$,
 with the distribution probability in eq.(\ref{2.16}). 
Using the replica method we obtain 
from  eq.(\ref{A.VIII}) and eq.(\ref{A.IX}):
\begin{equation}
\beta F=2\beta J_{k}\lambda^{\dagger}\lambda-\lim_{n\rightarrow 0}\frac{1}{Nn}[Z_{n}(SG)-1]
\label{A.XV}
\end{equation}
where $\alpha=1...n$ is the replica index and:
\begin{eqnarray}
Z_{n}(SG)=\langle\langle Z_{SG}^{n}\rangle\rangle &=& \int\displaystyle
\prod_{\alpha}^{n} D(\psi_{\alpha}^{\dagger}\psi_{\alpha})\exp{\{\displaystyle
\sum_{w\sigma}\sum_{i,j}g_{ij}^{-1}(w)\displaystyle
\sum_{\alpha}\psi_{i\sigma\alpha}^{\dagger}(w)\psi_{j\sigma\alpha}(w)\}}\nonumber\\          
& &\times\displaystyle\prod_{i,j}\langle\langle e^{\beta J_{ij}\displaystyle
\sum_{\alpha}S_{i\alpha}S_{j\alpha}}\rangle\rangle
\label{A.XVI}
\end{eqnarray}

The operators  $S_{i\alpha}$ are bilinear combinations of $\psi_{i\sigma\alpha}^{\dagger}(w)$, 
$\psi_{i\sigma\alpha}(w)$ from eq.(\ref{2.13}), then after performing the average in  
eq.(\ref{A.XVI}) we must use standard manipulations with gaussian identities [16,17] to 
linearize the exponent in eq.(\ref{A.XVI}). We obtain:
\begin{eqnarray}
Z_{n}(SG)=\int\displaystyle\prod_{\alpha,\beta}dq_{\alpha\beta}e^{-\frac{1}{2}(\beta J)^{2}N
\displaystyle\sum_{\alpha,\beta}q_{\alpha\beta}^{2}}\Lambda(\{q_{\alpha\beta}\})
\label{A.XVII}
\end{eqnarray}
where
\begin{eqnarray}
\Lambda(\{q_{\alpha\beta}\})&=&\int D(\psi_{\alpha}^{\dagger}\psi_{\alpha})
\exp{\{\displaystyle\sum_{i,j}\sum_{w,\sigma,\alpha}g_{ij}^{-1}(w)\psi_{i\sigma\alpha}^
{\dagger}(w)\psi_{j\sigma\alpha}(w)}+\nonumber\\
			    & &+\beta^{2}J^{2}\displaystyle\sum_{\alpha,\beta}q_{\alpha\beta}\sum_{i}S_{i\alpha}S_{j\alpha}\}
\label{A.XVIII}
\end{eqnarray}

We obtain for $Z_{n}(SG)$ at the replica symmetric saddle point:
\begin{eqnarray}
q_{\alpha\not=\beta}&=&q\nonumber\\
q_{\alpha\alpha}&=&q+\overline{\chi}
\label{A.XIX}
\end{eqnarray}
\begin{eqnarray}
Z_{n}(SG)\approx e^{-\frac{1}{2}(\beta J)^{2}N(\overline{\chi}^{2}+2q\overline{\chi})}
\Lambda(q,\overline{\chi})
\label{A.XX}
\end{eqnarray}
where
\begin{eqnarray}
\Lambda(q,\overline{\chi})=\displaystyle\int_{-\infty}^{\infty}\prod_{j} Dz_{j}
\prod_{\alpha}\displaystyle\int_{-\infty}^{\infty}\prod_{j} D \xi_{\alpha j}I_{\alpha}
(q,\overline{\chi},\{z_{j}\},\{ \xi_{\alpha j}\})
\label{A.XXI}
\end{eqnarray}
and
\begin{eqnarray}
I_{\alpha}(q,\overline{\chi},\{z_{j}\},\{\xi_{\alpha j}\})&=&\int 
D(\psi_{\alpha}^{\dagger}\psi_{\alpha})\exp{\{\displaystyle\sum_{w,\sigma}\sum_{i,j}
G_{ij\sigma
\alpha}^{-1}(w)\psi_{i\sigma\alpha}^{\dagger}(w)\psi_{j\sigma\alpha}(w)\}}\nonumber\\			    &=&e^{\displaystyle\sum_{w,\sigma}\ln{|G_{ij\sigma\alpha}^{-1}(w)|}}
\label{A.XXII},
\end{eqnarray}
where
\begin{eqnarray}
G_{ij\sigma\alpha}^{-1}(w)=g_{ij}^{-1}(w)-\delta_{ij}\sigma[\sqrt{2q}\beta Jz_{j}+
\sqrt{2\overline{\chi}}\beta J \xi_{\alpha j}]
\label{A.XXIII}.
\end{eqnarray}
              
Introducing Eq.(\ref{A.XXII}) and  Eq.(\ref{A.XXI}) in to Eq.(\ref{A.XX}), we obtain from  
Eq.(\ref{A.XV}) the expression for the free energy in  Eq.(\ref{2.15}) of the main text.       
%\end{document}

\newpage

\newpage

\section*{Figure captions}

{\bf Figure 1:}Phase diagram in the $T- J_{k}$ plane
as a function of $T/J$ and $J_{k}/J$ for fixed $J=0.05 D$, 
where D is the conduction bandwidth. The dotted line represents 
the "pure" Kondo temperature $T_{k}$.

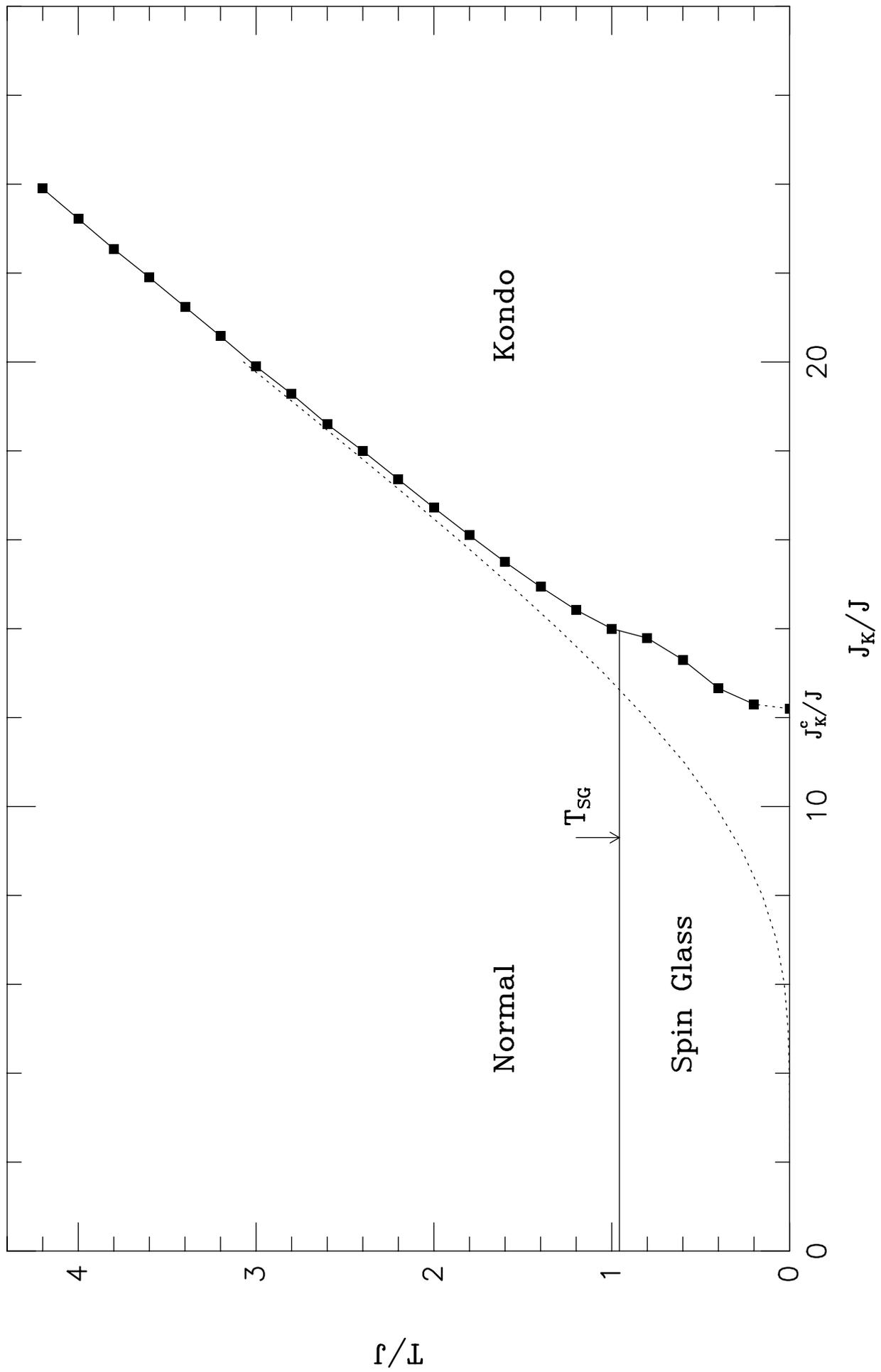
\end{document}